%% file: af_planet.tex
\documentclass[twocolumn]{aastex63}
\usepackage{tikz}
\usetikzlibrary{arrows.meta,
                calc, chains,
                quotes,
                positioning,
                shapes.geometric,
                shadows,
                backgrounds,
                fit,scopes}

\input{commands}

\hypersetup{linkcolor=red,citecolor=blue,filecolor=cyan,urlcolor=blue}
\usepackage[T1]{fontenc}
\usepackage{fontawesome}

\received{}
\revised{}
\accepted{}

\submitjournal{AJ}

\shorttitle{$\delta$~Sct exoplanets}
\shortauthors{Hey et al.}

\graphicspath{{./}{figures/}}

\begin{document}

\title{A search for transits among the $\delta$~Scuti variables in \textit{Kepler}}

\correspondingauthor{Daniel Hey}
\email{daniel.hey@sydney.edu.au}

\input{authors}

\newcommand{\candidates}{3}
\newcommand{\fpcount}{9}
\newcommand{\koidsct}{13}
\newcommand{\rev}[1]{\textcolor{black}{#1}}

\begin{abstract}
We search for transits around all known pulsating \dsct variables (6500 K < T$_{\rm eff}$ < 10\,000 K) in the long-cadence \kepler data after subtracting the pulsation signal through an automated routine. To achieve this, we devise a simple and computationally inexpensive method for distinguishing between low-frequency pulsations and transits in light curves. We find \candidates\ new candidate transit events that were previously hidden behind the pulsations, but caution that they are likely to be false positive events. We also examined the \kepler Objects of Interest catalog and identify \koidsct\ additional host stars which show \dsct pulsations. For each star in our sample, we use the non-detection of pulsation timing variations for a planet that is known to be transiting a \dsct variable to obtain both an upper limit on the mass of the planet and the expected radial velocity semi-amplitude of the host star. Simple injection tests of our pipeline imply 100\% recovery for planets of 0.5~R$_{\rm Jup}$ or greater. Extrapolating our number of \kepler \dsct stars, we expect 12 detectable planets above 0.5~R$_{\rm Jup}$ in \textit{TESS}. Our sample contains some of the hottest known transiting planets around evolved stars, and is the first complete sample of transits around \dsct variables. We make available our code and pulsation-subtracted light curves to facilitate further analysis. \href{https://github.com/danhey/dsct-exoplanet}{\color{black}\faGithub}


\end{abstract}

\section{Introduction} \label{sec:intro}

Planetary systems around the hot A/F stars are of considerable interest from the perspective of planetary formation, atmospheres, and habitability. The discovery of the `radius gap' in the distribution of \kepler planetary radii \citep{Fulton2017CaliforniaKepler,VanEylen2018asteroseismic} strongly supports the theory of photoevaporation of primary planetary atmospheres under the harsh light of their hot young stellar hosts \citep{Owen2013Kepler, Owen2017Evaporation}. Extremely highly irradiated planets around hot stars such as the A0 star KELT-9 \citep{Gaudi2017giant} have therefore been heavily studied as laboratories for understanding atmospheric loss generally, with evidence for significant exospheric envelopes of lost atmospheric gas \citep{Yan2018extended,Hoeijmakers2018Atomic}. Further examples of planets transiting hot stars are likely to attract similar interest.

Furthermore, while they are challenging for RV and photometry, many A/F stars are popular targets for direct imaging and astrometric planet searches. These include the A6 host $\beta$~Pictoris \citep{Lagrange2010Giant, Lagrange2020Unveiling}, with its edge-on debris disk, the F0 star HR~8799 \citep{Brandt2021First} and its four well-studied giant planets \citep{Marois2008Direct,Marois2010Images}; or the directly-imaged debris disks around the A0 \dsct Vega \citep{Matra2020Dust} and the A4 star Fomalhaut \citep{Kalas2005planetary}. Direct imaging is most-sensitive to planets and disk structures far from their hosts, and the inner architecture of hot star planetary systems is less-well understood.

The A/F stars' far-out habitable zone is easily accessible to astrometric planet detection \citep{Gould2003EarlyType}. While the habitability of such planets is expected to suffer from a harsh radiation environment \citep{Rugheimer2015UV} and short main-sequence lifespan \citep{Cuntz2016Exobiology}, it is not yet clear whether this is an insurmountable barrier to life \citep{Sato2014Habitability,Sato2017Climatological} and any such planets will be valuable to study. Planets around hot stars, whether in the habitable zone or otherwise, are valuable in exploring the architecture and habitability of these poorly-understood systems.

\rev{One class of A/F stars are the \dsct variables -- a class of intermediate-mass stars} that pulsate in pressure modes with frequencies between 5 and 100 d$^{-1}$ \citep{Guzik2021Highlights}. They exist along the classical instability strip across the HR diagram, typically between temperatures of 6,500 and 10,000~K \citep{Murphy2019Gaiaderived, Cunha2019Rotation}. The \dsct variables lack thick convective outer layers and thus retain much of their primordial angular momentum \citep{Cantiello2019Envelope}. As a result, they typically experience rotational velocities greater than 100~km/s \citep{Royer2009Rotation, Zorec2012Rotational} which heavily broadens their spectral lines, making it challenging to obtain precise RV measurements \citep{Ahlers2019Dealing}. For this reason, they are usually excluded from radial velocity exoplanet surveys which favor cooler, more slowly rotating host stars \citep{Howard2010CALIFORNIA, Grandjean2020HARPS,Rosenthal2021California}, although some searches directly target these stars (see, for example, \cite{Desort2007Planets, Borgniet2017Extrasolara}). Because of this difficulty in obtaining planetary precision RV measurements, few exoplanets are known, or even suspected, to exist around \dsct stars. Those known are typically rather massive hot Jupiters \citep{Galland2006Extrasolara, Herrero2011WASP33, Borgniet2014Extrasolar,Murphy2016Planet, Temple2017WASP167b, Martinez2020KELT25, Wong2020Exploring}.

While commonly seen as a nuisance, the pulsations in variable stars can be used to better constrain the evolutionary state of the system.  For example, in solar-like oscillators, global asteroseismic parameters have been used to better characterize the system under a joint asteroseismic and transit analysis (see, for example, \citealt{Gilliland2010ASTEROSEISMOLOGY, Huber2018Synergies, Chontos2020TESS}). For \dsct variables, the splitting of pulsation modes can reveal the stellar rotation rate and the plane of obliquity \citep{Kurtz2014Asteroseismic, Chen2017Rotational}. Unfortunately, the rapid rotation of \dsct variables not only spoils their absorption line features, but also causes large and uneven rotational splittings in the excited modes of oscillation, such that multiplets of adjacent modes can overlap \citep{Reese2009Pulsation, Mirouh2019Mode}. In addition, the asteroseismic inference from \dsct light curves has lagged behind many other types of pulsating variables until recently \citep{Bedding2020Very}, making precise stellar parameters for \dsct stars difficult to obtain. Now, however, some \dsct stars have ultra-precise parameters derived from asteroseismology \citep{Murphy2021precise}, opening them up as a powerful diagnostic as exoplanet host stars.


With the advent of extremely high-precision space-based telescopes such as \textit{Kepler}, \textit{TESS} and the upcoming \textit{PLATO} mission \citep{Koch2010Kepler,Borucki2010Kepler, Ricker2014Transiting, Rauer2014PLATO}, a large influx of light curves for \dsct stars are already available. It is estimated that approximately half of the stars located within the classical instability strip are pulsating above the $\sim$10~ppm detection threshold provided by \textit{Kepler} \citep{Murphy2019Gaiaderived}. 

The photometric amplitude of the stellar pulsations in \dsct variables may be on the order of 10 to 100,000~ppm. In comparison, the decrease in stellar flux from a transit of an Earth-like planet is on the order of 100~ppm. Transits whose amplitudes are bigger than the pulsation amplitudes are relatively easy to spot. Difficulties arise when the pulsational variability exceeds that of the transits, where it can be said that such transits are ‘hidden’ in the pulsations. To find these hidden transit signals, the light curve should be processed in a manner that removes the pulsation signal (referred to as `cleaning' in this paper) but preserves the transit. This method was used by \cite{Zieba2019Transiting} to find transiting exocomets around the \dsct star $\beta$~Pictoris. We are fortunate in the fact that most transits have clearly non-sinusoidal shapes. On the other hand, pulsations in \dsct variables behave highly sinusoidally. This lets them be easily expressed as the sum of sine waves that can be subtracted (or divided out) from the light curves, allowing for transit searches in the residuals. In frequency space, the transits appear as peaks at integer multiples of the orbital frequency as a result of their rectangular shape.

\citet{Sowicka2017Search} conducted such a search for transits around pulsating stars in \kepler by first modeling the stellar pulsations as sinusoids and then subtracting this model. 
This search was limited to stars with short-cadence observations and temperatures $6000 \text{K} < T < 8500~\text{K}$, and found two candidates, around KIC~5613330 and KIC~8197761; the latter of which was confirmed by RV follow-up to be an eclipsing binary. 

In this paper, we seek to extend their search and characterize the occurrence rate of exoplanets around the \dsct variables in the entire \kepler long-cadence field. Taking a similar approach to \citet{Sowicka2017Search}, we design a pipeline that iteratively subtracts the \dsct pulsations to facilitate a transit search, and introduce a simple method for distinguishing between pulsations and transits in the low-frequency regime. We then search for transit events in the cleaned light curves through a combination of manual inspection of Box Least Squares (BLS) periodograms and a semi-automated transit search with Transit Least Squares (TLS) in Section~\ref{sec:pipeline}. We identify \candidates\ previously unknown candidate transit events, and fit their planetary parameters with a transit model in Section~\ref{sec:results}. We then search the light curves of the candidate and confirmed planets in the \textit{Kepler} objects of interest table for \dsct oscillations, completing our sample of \dsct host stars. Finally in Section~\ref{sec:pulsation}, we exploit the timing variations of the stellar pulsations over the orbit to provide a robust upper limit on the planetary mass, and consequently, an upper limit on the radial velocity precision required to rule them out as candidates or confirm their planetary status. We make available our code and pulsation subtracted light curves\footnote{\url{https://github.com/danhey/dsct-exoplanet}}.

\section{Observational data} \label{sec:data}

\begin{figure}
    \centering
    \includegraphics{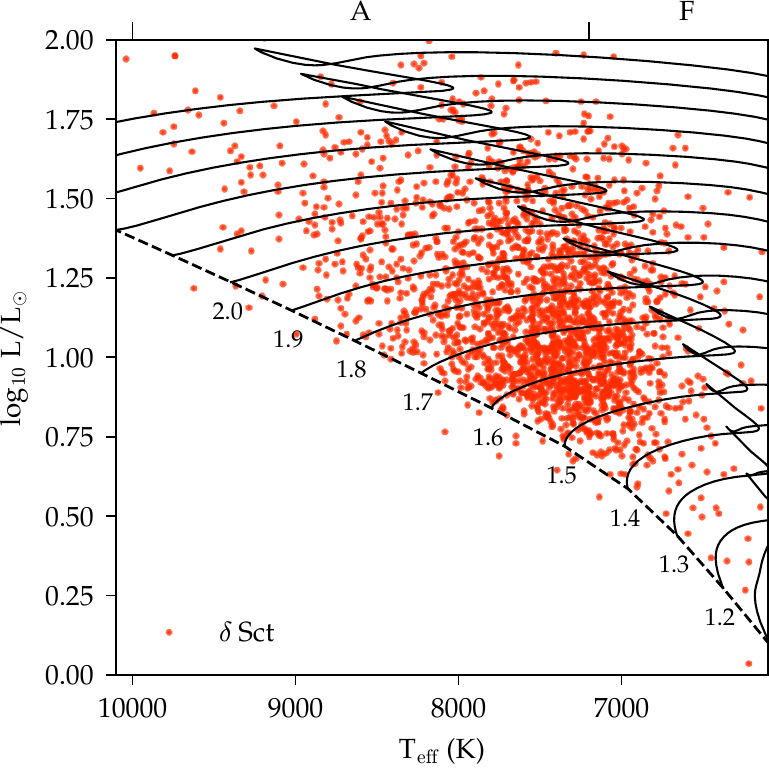}
    \caption{HR diagram of the \dsct sample. We make no distinction for hybrid pulsators -- those which oscillate simultaneously in p- and g-modes. \rev{The evolutionary tracks were calculated in MESA v10108 \citep{Paxton2011Modules} with $X = 0.71$ and $Z = 0.01$. See section 4.2 of \citet{Murphy2019Gaiaderived} for a full description of the tracks.}}
    \label{fig:HRD}
\end{figure}

We chose our sample to be the known \dsct variables in the \kepler field for three reasons. Firstly, most \dsct stars in the \kepler field have been identified. Secondly, their masses, temperatures, and other fundamental parameters have been well-characterized over the past several years \citep{Murphy2019Gaiaderived, Berger2020GaiaKepler, Guzik2021Highlights}. Finally, the 4-year time-span of \kepler data allows pulsation timing variation studies, whose sensitivity scales with the SNR of the pulsations \citep[see, for example,][]{Murphy2014Finding}. Despite our focus only on the \kepler \dsct stars, we note that all of the principles in this paper are readily applicable to other space based photometric surveys, including \textit{TESS}.

We compiled a catalog of \kepler \dsct stars by combining the \dsct catalog of \cite{Murphy2019Gaiaderived} with those of \cite{Balona2018Gaia} and \cite{Bradley2015Results}. We excluded the slowly pulsating B-stars and perhaps mythical `Maia' variables \citep{White2017Kepler} cataloged by \cite{Balona2018Gaia}. We also chose not to include the rapidly oscillating Ap stars, since only few are known and they have been intensely scrutinized already \citep{Holdsworth2018Suppressed, Hey2019Six}. After merging the catalogs, we were left with 2354 stars with \dsct pulsations, some of which are hybrid pulsators, which pulsate in both high-frequency p-modes and low-frequency gravity (g) modes simultaneously, typically known as \dsct / $\gamma$ Doradus hybrids \citep{Grigahcene2010Kepler, Uytterhoeven2011Kepler}.

In order to search for transits, we used the light curves from the 4-year nominal \textit{Kepler} mission. \textit{Kepler} was a space-based telescope that collected precise photometric data for around 150,000 stars simultaneously in a 115~${\deg}^2$ field of view. The core science goal of the \textit{Kepler} mission was to detect Earth-like planets in the habitable zone \citep{Borucki2010Kepler, Koch2010Kepler}. A secondary goal of this mission was asteroseismology \citep{Gilliland2011Kepler}: with a photometric precision verging on the $\mu$mag level, \textit{Kepler} provided unprecedented insight into stellar pulsations. 

The \textit{Kepler} data are available in two observing modes: long cadence (LC) and short cadence (SC). The LC data are integrated over 29.4-min observations, whereas the SC data are integrated over 1-min observations, with up to 512 stars being allocated to SC at any given time on a priority basis. Both sets of data were stored on-board the spacecraft and regularly down-linked to Earth every 32~d, along with a periodic roll to re-orient the spacecraft's solar panels, introducing semi-regular gaps in the data. The \textit{Kepler} data are thus organized into quarters, which we downloaded from the Mikulski Archive for Space Telescopes (MAST) repository. Light curves have both simple aperture photometry (SAP) and pre-search data conditioning simple aperture photometry (PDCSAP; \citealt{Twicken2010Presearch, Smith2012KeplerPresearch}). We chose to use PDCSAP v9.3, which applies co-trending basis vectors to the SAP flux to remove long-term and systematic trends, which is effective for both pulsations and transits.

\subsection{Stellar properties}

We took temperatures and luminosities for our sample from \cite{Berger2020GaiaKepler}. For stars not in this catalogue, we calculated the luminosity following the same method which we briefly outline now. Reddening was obtained from the 3D maps provided by \cite{Green2018Galactic}, and distances were calculated using the Bayesian approach described by \cite{Bailer-Jones2018Estimating} with \textit{Gaia} eDR3 parallaxes. We did not apply an offset in the parallax, since \cite{Murphy2019Gaiaderived} found it results in underestimated luminosity for stars in this temperature range. We used the \textit{Gaia} G-band magnitudes, and bolometric corrections from the \textit{MIST} isochrones \citep{Dotter2016MESA}. We show the location of the stars on the HR diagram in Fig.~\ref{fig:HRD}. 

\section{The pipeline} \label{sec:pipeline}

\begin{figure}
    \centering
    \includegraphics[width=\linewidth]{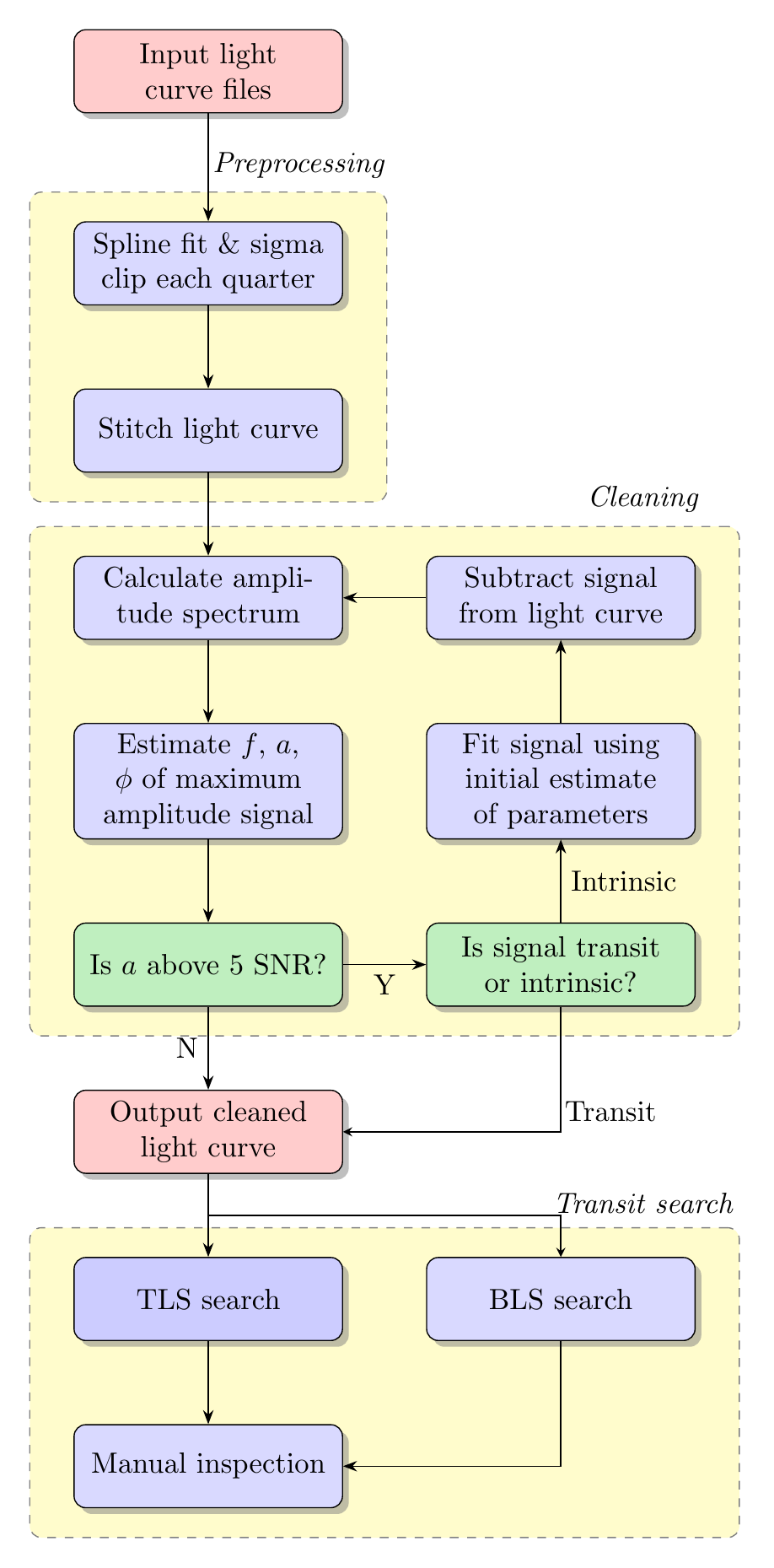}
    \caption{Flowchart of the main functions of the pipeline.}
    \label{fig:pipeline}
\end{figure}

We designed a pipeline to search for signals hidden by the pulsations of the \dsct stars. The pipeline has multiple components which perform the following operations in order: light curve corrections, iterative fitting and subtraction of pulsation modes, and a transit search. We describe each aspect of the pipeline in detail below, and provide a flowchart of its main operations in Fig.~\ref{fig:pipeline}.

\subsection{Light curve corrections}

In the first phase of the pipeline, we processed the individual \kepler quarters to remove slow variations for each star. To do this, we performed a 3-sigma clipping of the light curves and subtracted a 1D spline fit of 2nd degree made using the \textsc{Scipy} package \citep{Virtanen2019SciPy}. The purpose of the spline fit was to remove systematic variations which were unaffected by the PDCSAP pipeline, in particular, variations that result in extremely non-sinusoidal low-frequency signals in the final light curve. We chose a $2^{\rm nd}$ order spline so any potential transits would not be removed. We then stitched each quarters into a final combined light curve.

\subsection{Removing pulsations}

\begin{figure*}
    \centering
    \includegraphics{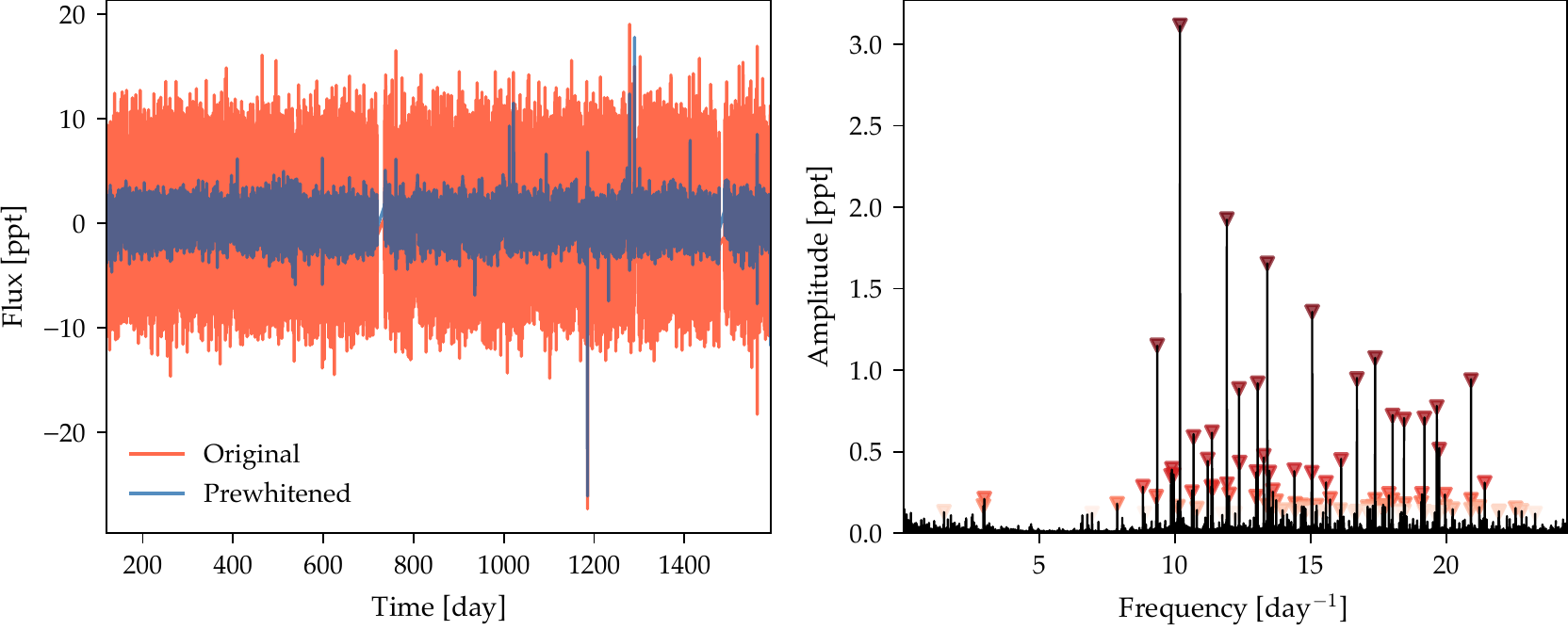}
    \caption{Example of light curve cleaning for the known planet hosting \dsct variable KIC~5202905. Left panel: the original flux (red) and after cleaning away the pulsations (blue). Right panel: The amplitude spectrum of the original flux. Over-plotted are the fitted frequencies -- the opacity of the points are proportional to their amplitude. Note that the planetary signal is almost undetectable at this scale.}
    \label{fig:clean_example}
\end{figure*}

The cleaning routine is an iterative procedure that is widely used for pulsating stars (also known as prewhitening, \citealt{Lenz2004Period04}). The method fits and subtracts signals in the time domain of the form

\begin{equation}
    y(t) = A\cos{(\omega t + \phi)},
    \label{eqn:model}
\end{equation}

\noindent where $A$, $\omega$, and $\phi$ are the amplitude, angular frequency, and phase of the oscillation mode respectively. The routine selects the frequency peak in the amplitude spectrum of highest amplitude at each step, calculates an initial estimate of the amplitude and phase, and then fits Eq.~\ref{eqn:model} to the light curve. The resulting fit is subtracted from the total flux for subsequent iterations. The iteration is continued until either one of two conditions is satisfied: there is no signal remaining above a signal-to-noise ratio (SNR) of 5, or more than 100 signals have been removed. We calculated the SNR by dividing the amplitude spectrum by an estimate of the background noise, calculated through a moving median filter. The pipeline only cleans between the frequencies of 1 to 47~d$^{-1}$. \rev{The upper limit of 47~d$^{-1}$ was chosen as it is approximately $f_{s} - 1$, where $f_{s}$ is the sampling frequency of the \kepler\ LC data. \cite{Murphy2019Gaiaderived} found that 17.9\% of \dsct stars have their strongest pulsation frequency above the Nyquist limit ($0.5f_s$), and so we consider our chosen limit to be acceptable. Even if the pipeline were to clean an aliased signal, the true signal would also be removed in the process. We note that calculation of the amplitude spectrum is the most computationally intensive part of the pipeline, so our frequency choice is a trade-off between accuracy and speed.} We discuss low-frequency signals in the next section.

All of the parameters of Eq.~\ref{eqn:model} can be well determined and used as starting points before fitting. Prior to fitting to Eq.~\ref{eqn:model}, we calculated an initial estimate of the frequency, amplitude, and phase. The frequency of maximum amplitude was calculated from a three-point parabolic interpolation routine applied to the periodogram. The amplitude was then calculated from the periodogram at this single frequency estimate. Finally, the phase was obtained from the real ($\Re$) and imaginary ($\Im$) components of the Discrete Fourier Transform ($z$) at the given frequency,

\begin{equation}
    \phi = \tan{\Big(\frac{\Im{z}}{\Re{z}}\Big)}^{-1}.
\end{equation}

To fit Eq.~\ref{eqn:model} we used a non-linear least squares algorithm as implemented in Scipy \citep{Virtanen2019SciPy}. We calculated the gradient function of Eq.~\ref{eqn:model} as a faster alternative to numerical estimates required by the solver,

\begin{eqnarray}
\nabla y = \left( \begin{array}{c}
    -At\sin{(\omega t + \phi)} \\
    \cos{(\omega t + \phi)} \\
    -A\sin{(\omega t + \phi)}
	\end{array}\right).
\label{eq:Dj}
\end{eqnarray}

We note that in most cases, actual optimization of the parameters is unnecessary. Indeed, the initial estimate of the parameters is usually sufficiently close to the true value to the point that optimization completes within a few iterations, and the optimized value is functionally identical to the initial guess. Despite this, the optimization is cheap to perform (<1~s per pulsation mode) and thus included for the sake of completeness.

The iterative aspect of the cleaning is key to the light curve processing. It is not feasible to select, fit, and subtract all oscillations lying above a given SNR simultaneously from the light curve, because most \dsct stars exhibit amplitude modulation \citep{Bowman2016Amplitude}. Amplitude modulation causes the mode amplitude to vary over the the length of the \kepler data. As a result, it sometimes takes several or more iterations before a single oscillation mode is subtracted.
We show pre-whitening results for the known \dsct planet host, KIC~5202905 in Fig.~\ref{fig:clean_example}. Note that the cleaning halts when it reaches the low-frequency regime, where the transit appears. 

\subsection{Dealing with low-frequency variability}

A major difficulty in removing pulsations from \dsct light curves is that they commonly exhibit low-frequency variations, caused either by spot-based rotational modulation or low frequency gravity (g) modes \citep{Breger2011Rotational, VanReeth2015Gravitymode, VanReeth2018Sensitivity}. Attempting to clean at low frequencies requires the user to make a choice between whether the observed signal is intrinsic to the star or instead caused by a transit. Here, we devise a simple method for differentiating between transits and intrinsic stellar variability through the use of the Bayesian Information Criterion \citep[BIC;][]{Schwarz1978Estimating, Neath2012Bayesian}. The BIC is a type of model selection that calculates a quantitative value that can be used to determine which model best represents the data:
\begin{equation}
    {\rm BIC} = k \ln{n} - 2 \ln{\hat{L}},
\end{equation}
where $k$ is the number of parameters in the model, $n$ is the number of samples, and $\hat{L}$ is the likelihood of the model calculated at the maximum a posteriori (MAP). That is, the value of the likelihood at the optimized model parameter values. 

To implement the BIC in our search, we modelled intrinsic variability (pulsation or rotation) following Eq.~\ref{eqn:model} and modelled transits using a Box Least Squares (BLS) model. For each signal below 1~d$^{-1}$ with a SNR greater than 5, we followed the iterative cleaning procedure prescribed above. However, at each step of cleaning, the transit model was also fitted to the observed signal and the BIC was calculated. If the BIC favored the pulsation model then the signal was removed and the process repeated. On the other hand, if instead the BIC favored the transit model, then the procedure was halted and the pipeline moved onto the next star.

To test the efficiency of the BIC as a discriminator between transits and intrinsic variability, we ran it on 1000 randomly selected light curves from our \kepler sample. For each light curve, we injected a transit calculated using the \textsc{batman} Python package \citep{Kreidberg2015batman}, with a random orbital period between 0.5 and 50~d, and planetary radius from 0.01 to 0.5~R$_\odot$, both drawn from uniform distributions. For each signal to be cleaned, we calculated the BIC for both a transit and pulsation model fit. We recorded the true positive, true negative, false positive, and false negative rate. For example, if the algorithm identified the signal as a transit at the correct orbital period (within a tolerance of 0.001~d), it counted one true positive. Similarly, if the algorithm identified the signal as a transit and it was actually a pulsation, we counted one false positive.

\begin{figure}
    \centering
    \includegraphics{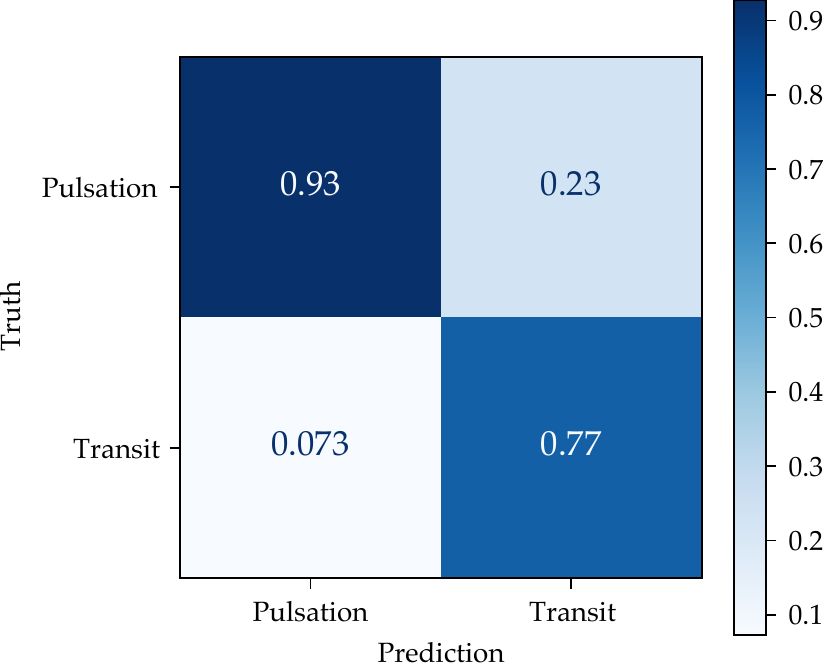}
    \caption{Confusion matrix of injection testing using the BIC to differentiate between transits and intrinsic variability in a light curve. The BIC has been normalized to 1.}
    \label{fig:BIC}
\end{figure}

We show the results of our injection testing as a confusion matrix in Fig.~\ref{fig:BIC}. We note that the algorithm is heavily biased towards classifying variability as intrinsic -- because the cleaning will continue if a pulsation is observed, whereas the cleaning stops if a transit is detected. As a result, there are many more instances of pulsation true positives than transits in the simulation. In utilizing the BIC to clean only pulsations, we seek to maximize our recall score at the cost of precision, so we do not accidentally mis-classify a transit as a pulsation. We see from our results that our recall is effectively maximized: the number of transits accidentally classified as pulsations is around 7\%, although 23\% of pulsations were mis-classified as transits.

\subsection{Transit search}
\label{transit_search}

Once the pulsations and intrinsic variability were removed, we then searched each light curve for transiting planets.
For each quarter of \kepler\ data, we removed variability on timescales significantly longer than transit durations with the Savitsky-Golay filter implemented in the \texttt{lightkurve} package \citep{GeertBarentsen2019KeplerGO}.
We filtered using a second-order polynomial and a window length of 201 cadences (4.1 days), and we treated independently any light curve segments separated by a gap larger than 11 cadences (5.5 hours), typically the result of monthly data downlinks. We then re-stitched the light curve into a single data set.

We searched for transiting planets with the \texttt{transitleastsquares} (TLS) algorithm of \cite{Hippke2019Optimized}. 
TLS broadly follows the approach of the Box Least Squares (BLS) algorithm \citep{Kovacs2002boxfitting}, updated to search for periodic signals with realistic transit shapes and appropriate limb-darkening rather than a simple inverse tophat.
We restricted our search to periods between 0.5 and 50.0 days; 77\% of all \kepler\ objects of interest fall in that period range.
Given the large radii of the stars in our sample, we expect planets with periods shorter than 0.5 days to be quickly tidally engulfed \citep[e.g.][]{Li2014DYNAMICS, Patra2017Apparently} and longer period transiting planets to be below the threshold for detectability. We applied an oversampling factor of 0.2, which results in a search over 10,655 periods over this range. We also performed a complementary search using the Box Least Squares periodogram, with the orbital period restricted instead to between 0.5 and 100 days.

For each system, we recorded the signal detection efficiency as a function of period, the binned and un-binned phase-folded flux at the time of the most significant signal and offset by a phase of 0.1, and summary statistics on the best-fitting orbital period, transit duration, and transit depth. We then visually inspected each of these outputs for all systems to identify candidates. 

To quantify the efficiency of our pulsation removal and transit detection pipeline, we performed a series of basic injection-and-recovery tests. We generated a series of synthetic transiting planets using the \textsc{batman} package \citep{Kreidberg2015batman} by varying their radii from 0.1 to 5~$_{\rm Jup}$ at a fixed orbital period of 10~days. These were then injected into the original PDCSAP light curves. For simplicity, we set the other transit parameters to fixed values. We assumed circular orbits ($e=0$), and a radius for the host star of $1.7~R_\odot$, a typical value for an A/F-type star.

After running the injected light curves through our pipeline, we recovered 100\% of planets larger than around 0.5~R$_{\rm Jup}$, around 80\% with 0.36~R$_{\rm Jup}$, and 50\% at 0.25~R$_{\rm Jup}$. We only detected the smallest radius (1~R$_{\rm Jup}$) planet in one of our systems. These tests give us confidence that our method is sensitive to, at the least, the 0.5~R$_{\rm Jup}$ planets.








\section{Results} \label{sec:results}


\input{tables/results.tex}
We identified 32 possible transiting exoplanet candidates from our cleaned light curves. We cross-matched this sample against the KOI catalogue and found that 13 are listed  as \textit{False Positives}, two are \textit{Confirmed}, two are \textit{Candidates}, and the remaining 16 are not in the catalogue.

For some of our candidate transits, the dips did not appear as a typical transit-like profile (with a U-shaped transit due to limb-darkening). However, as transiting systems exist whose geometry does not generate flat minima in the light curves, we decided to retain them in our sample. A number of effects can change the transit shape, e.g. gravity darkening of a rapidly rotating, oblate host star can cause asymmetric transit shapes \citep{Barnes2009TRANSIT}. We caution that some of these events may not be real transits, but could rather be due to imperfectly removed pulsations mimicking a transit-like light-curve shape. As such, we do not claim that the candidates here show true transits, but rather that they are worthy of more in-depth study.

\begin{figure}
    \centering
    \includegraphics{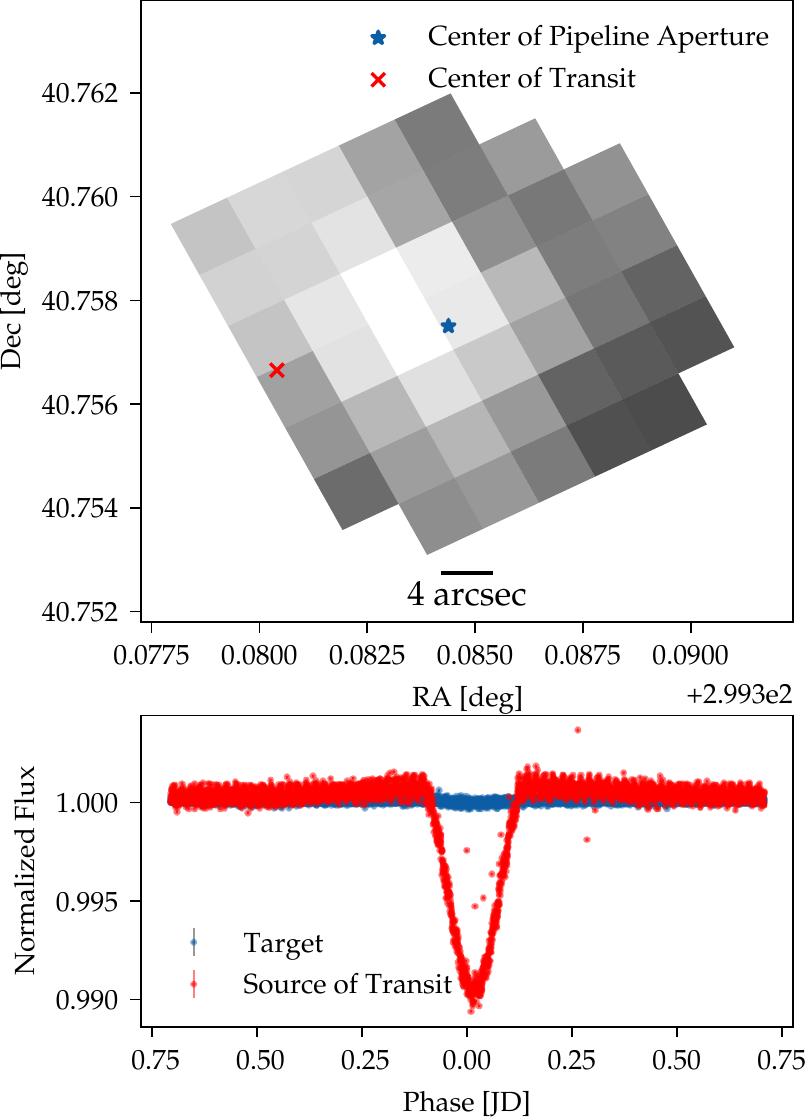}
    \caption{Contamination for KIC~5565497, a known False Positive from the Kepler Objects of Interest catalog. \textbf{Top:} the target pixel file with the locations of the center of pipeline aperture and transit source marked. The transit is several arcseconds away from the center of the pipeline aperture and is likely to originate from a nearby eclipsing binary. \textbf{Bottom:} Light curves created on the pipeline (blue) and transit (red). The pipeline aperture light curve is constant.}
    \label{fig:contaminante}
\end{figure}

We checked each candidate for obvious signs of blending and nonphysical scenarios, using the \textsc{contaminante} software \footnote{\url{https://github.com/christinahedges/contaminante}}, which models the source of the transit signal in relation to the center of the aperture pipeline in the target pixel file (TPF). Stars with deviations exceeding one arcsecond between the center of transit and aperture on the TPF are likely to be visual blends caused by a background eclipsing binary. We show an example of an obvious background eclipsing binary system, KIC~5565497, in Fig.~\ref{fig:contaminante}. We note that for 15 of our stars, we were not able to isolate the transit on the TPF. This was due to the fact that most of the transits are hidden behind the pulsations, and so can not be accurately modeled until the pulsations are removed, which was not done on a TPF aperture. 

We also checked whether the density of the star implied from the transit agreed with independent calculations from our sample. From \citet{Seager2003Unique, Sandford2017Know}, it is known that the stellar density ($\rho_*$) can be measured from the transit period ($P$) and duration ($T$),
\begin{equation}
    \rho_* = \frac{3P}{\pi^2 G T^3},
    \label{eq:density}
\end{equation}
where $G$ is the gravitational constant. We do not exclude candidates from our sample if the implied stellar density is significantly different from our calculations, because those calculations assume a perfect sphere, which is not the case for the typically oblate \dsct variables. Likewise, there remains a possibility of the transit occurring on another nearby star in a binary system, where the \dsct pulsations are contaminating the light curve. Regardless, we calculate the implied stellar density for all of our stars and used them to inform our vetting process.

\begin{figure*}
    \centering
    \includegraphics{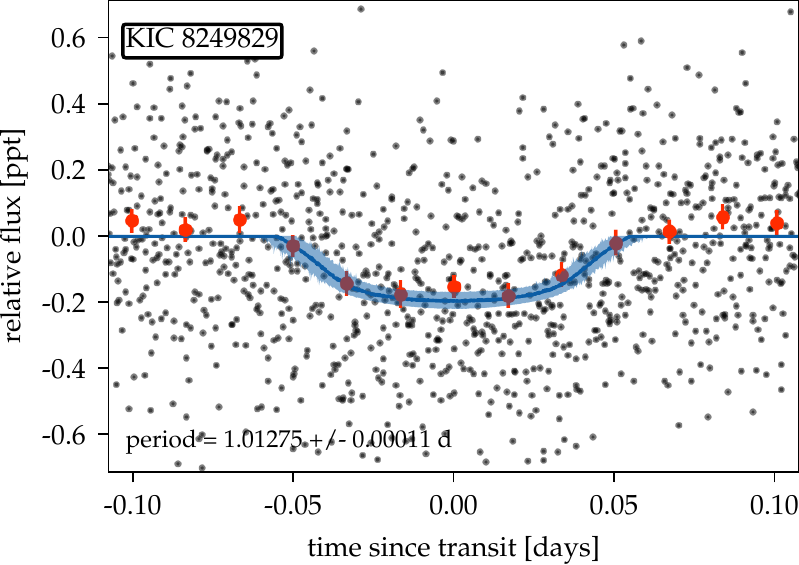}
    \includegraphics{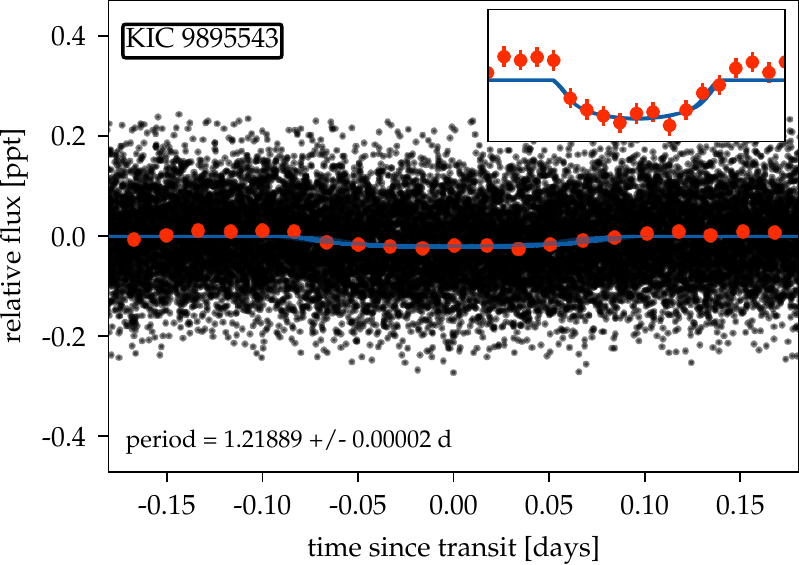} \\
    \includegraphics{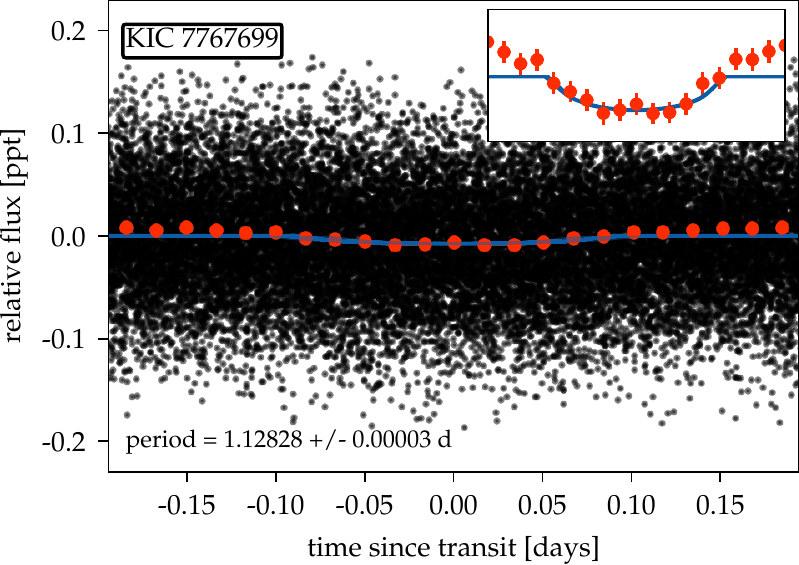}
    
    
    \caption{Transit fits for the \candidates\ candidates in our sample. We plot the flux in black points, the binned flux in red points, and plot the median and standard distribution of the posterior light curve fit as a blue line.}
    \label{fig:planet_fit}
\end{figure*}

For each candidate that was not listed as a KOI, we fitted a transit model to the light curve using the exoplanet \textsc{Python} package, written on top of \textsc{PyMC3} \citep{Foreman-Mackey2021exoplanet}. This model consisted of both a standard transit model and a Gaussian Process noise model with a simple harmonic oscillator kernel. For the orbit, we parametrized over duration and assumed a circular orbit, after preliminary models indicated that the eccentricity is poorly constrained.

\subsection{\textit{KOI} \dsct variables}

In addition to our candidate sample of known \dsct stars, we also searched for \dsct oscillations in all KOIs with a temperature between 6000 and 10,000~K, roughly corresponding to the classical instability strip for \dsct pulsators. We considered both `confirmed' and `candidate' planet designations. \dsct stars were then identified by calculating the skewness of the amplitude spectrum above 10~d$^{-1}$, following \cite{Murphy2019Gaiaderived} and \cite{Bedding2020Very}. Stars with a skewness less than unity are typically non-pulsating and were removed. The remaining stars were visually inspected to identify \dsct pulsations. We note that this sample does not include the single planet discovered around a \dsct variable by pulsation timing \citep{Murphy2016Planet}, because the KOI catalog only lists transiting planets. We summarize the key results of the fitting and KOI \dsct variables found in Table~\ref{tab:planet}. \rev{We report the observing cadence, the transit fit parameters; the orbital period, time of transit ($T_0$), the duration of the transit, the fitted radius of the planet, the ratio of the semi-major axis to the stellar radius (a/R$_*$), and the ratio of implied to calculated stellar densities from Eq.~\ref{eq:density}.} We use the KOI values for stars that were already known.

In our search for \dsct variables in the KOI, we found 13 stars that show both transits and \dsct pulsations out of a total 1171 objects of interest which we list in Table~\ref{tab:planet}. Ideally, all of these systems would have appeared in our cleaning transit search. However, only four systems from the KOI were found in our transit search. We attribute this to two causes: our \dsct sample is incomplete, and the transit search algorithm is not 100\% effective. The largest contributor to the \dsct sample was based on a \mbox{6,500 - 10,000~K} temperature cut \citep{Murphy2019Gaiaderived}, whereas the KOI search was performed between 6,000 - 10,000 K. This excluded three of the KOI sample from our \dsct sample: KIC~3964109, KIC~9111849, and KIC~11013201. Additionally, our search was only performed on the \textit{Kepler} LC data. From Table~\ref{tab:planet}, seven of the 13 stars were observed in LC mode. Of those seven, six were in our \dsct planet search sample, and we recovered four transit events. The two systems that we did not identify in our search were KIC~5617529 and KIC~6116172. Although KIC~6116172 is a candidate for a 3 planet system, it is not surprising that it did not appear in our search because strongest transit signal in the light curve is at an orbital period of 111 days, which exceeds the limits of our search (1 to 100 days). The remaining sample in the KOI were all observed in SC mode, which our \dsct sample did not use.

\subsection{Candidates}

From our 16 new candidates from the transit search, only three transit models converged, of which only KIC~8249289 shows the typical flat-bottom minimum expected of a transit. The other two show an extremely long duration and either a shallow transit, or a deep V-shaped eclipse, implying that the transit geometry is grazing and/or it is a background EB (Fig.~\ref{fig:planet_fit}). We are unable to confirm or rule out these candidates from their transit shapes alone, and we caution that it is likely they are all false positives. This is supported by the fact that the binned flux outside the transit appears to vary sinusoidally, which is indicative of an incorrectly removed pulsation or data processing error. We inspected the amplitude spectrum of these three candidates and found no significant pulsation structure around the orbital frequency. In saying that however, the shallow, long duration transits in KIC~9895543 and KIC~7767699 are not unusual: the candidate planetary system KIC~6116172 from the KOI also shows a similar transit profile. For all three candidates, although they lie close to the boundary of their host star (a/R$_*\sim$1), their implied stellar density from the transit duration is somewhat close to the calculated value.

As a final test of the candidates, we performed ephemeris matching by cross-matching our candidates against the entire KOI catalog \citep{Coughlin2014Contamination}. We checked the period and time of transit ($T_0$), and found no matching objects within a period tolerance of 0.01~days and $T_0$ tolerance of 1~day. We note however that the \rev{radii} of the candidate planets are much lower than our expected limits of recovery from the injection tests. 

\subsection{False positives}

\input{tables/fp}

Our transit search yielded 28 false positives. As mentioned, 13 of them are previously known false positives from the KOI catalog, and the remaining 15 were either found to be contaminated, or the transit fit failed to converge. We summarize the new false positives and their reason for exclusion in Table~\ref{tab:fp}. Of these false positives, KIC~4380834, KIC~5724523 \rev{and KIC~9471419} have a nonphysical ratio between their semi-major axis and radius of their host star (a/R$_*$ << 1). Their values imply an orbital position within the envelope of their \dsct host, and thus, were downgraded to false positives. 



\section{Pulsation timing variations} \label{sec:pulsation}

Hot stars, which include \dsct stars, are not amenable to radial velocity surveys for planetary sized companions because of their rapid rotation. However, their highly coherent pulsations allow for the study of timing variations in a manner similar to eclipse timing variations and pulsar timing \citep{Murphy2014Finding, Murphy2015Deriving}. In a multiple system, the pulsating stars' orbit around the barycenter leads to a change in path length for starlight traveling to Earth. If at least one star is pulsating, the path length changes manifest as variations in the light arrival time of the pulsations. The minimum detectable mass of a companion found through pulsation timing relies on the signal-to-noise ratio (SNR) of the pulsation, and its coherency over the orbital phase. These values are independent of the rotation of the host star. Previously, it has been shown that pulsation timing can be applied to \kepler\ objects of interest to recover a signal \citep{Balona2014Binary}. However, information about the system can still be gained even when no signal is observed. That is, the upper limit of the mass of the companion can be found.

For an exoplanet of sufficient mass, mutual gravitation between the planet and its pulsating host star leads to variations in the arrival times of the stellar pulsations. These variations manifest as time-dependent phase changes in the pulsations modes. Orbital phase changes are distinguished from frequency variations intrinsic to the star by the fact that they must affect all pulsation modes equally.

The sensitivity of pulsation timing depends on the quality of the celestial `clock' used \citep{Compton2016Binary}. The milli-second level precision of pulsars has been used to great success for identifying exoplanets through their light travel time variations \citep{Wolszczan1992Planetary, Sigurdsson2003Young,Suleymanova2014Detection, Starovoit2017Existence}. On the other end of the scale, applications of pulsation timing towards variable stars oscillating at lower frequencies, such as the A/F-type $\delta$ Scuti stars, has been highly successful in identifying binary systems \citep{Murphy2018Finding, Murphy2020Finding}. The sensitivity of pulsation timing is proportional to the oscillation frequency, and, as a result, only one planet has successfully been detected using pulsation timing with \dsct stars \citep{Murphy2016Planet}. We recently introduced a new approach which simultaneously fits the phase variations across the orbit of a star using a forward model \citep{Hey2020Forward}. This method was shown to be more sensitive and robust than previous approaches for identifying planets with pulsation timing.

The orbital period obtained from a transiting exoplanet can be used to constrain a model of pulsation timing variations. We can use the pulsation timing variations as a proxy for radial velocities of exoplanets transiting \dsct stars. Even if no timing variation is observed in the pulsations, the absence of signal can be used to place an upper limit on the mass of the companion. We performed this analysis on all 13 KOIs with $\delta$ Scuti pulsations, and our three new candidates to constrain the planetary mass function. To the best of our knowledge, this is the first attempt to vet transiting exoplanet candidates from timing of their host star oscillations.

\input{tables/pm_dists}

We briefly describe the theory of pulsation timing below. For a full explanation, see \cite{Murphy2014Finding, Hey2020Forward}. The observed relative flux $y(t)$ of a pulsating star with $J$ modes undergoing mutual gravitation with a companion around a common barycenter varies according to
\begin{equation}
y(t) = \sum_{j=1}^J  A_j \cos (\omega_j[t-  \tau] + \phi_j),
\label{eq:luminosity}
\end{equation}
where $A_j$ is the amplitude of mode $j$, $\omega_j = 2\pi\nu_j$ is its angular frequency, and $\phi_j$ is its phase. $\tau$ acts as an additional phase term, and represents the time delay -- the projected light arrival time of the pulsations from the star -- and is described by the same orbital elements as radial velocities:
\begin{equation}
    \tau = -\frac{a \sin{i}}{c} \frac{1 - e^2}{1 + e \cos{f}} \sin{(f + \varpi)},
\end{equation}
in units of seconds, corresponding to a change in path length measured in light-seconds. We follow the conventions that $a_1 \sin i$ denotes the projected semi-major axis of the primary, $e$ is the eccentricity, $\varpi$ is the angle between the ascending node and the periapsis, and $c$ is the speed of light. These conventions are chosen so that when the star is on the far side of the barycenter with respect to Earth, the time delay will be at its (positive) maximum.

In previous papers, the time-delay has been modeled for binary systems where the orbital period was not known a priori \citep{Murphy2020Finding, Hey2020Forward}. In those cases, the orbital period was obtained by either subdividing the light curve into equal segments or by `brute-forcing' a model over a grid of orbital periods to observe where the model likelihood peaks. For the stars in this paper, however, the orbital period is extremely well constrained from the observed transit signal. This allows us to fix the pulsation timing model at the exact orbital period and fit the remaining parameters. 

\begin{figure}
    \centering
    \includegraphics{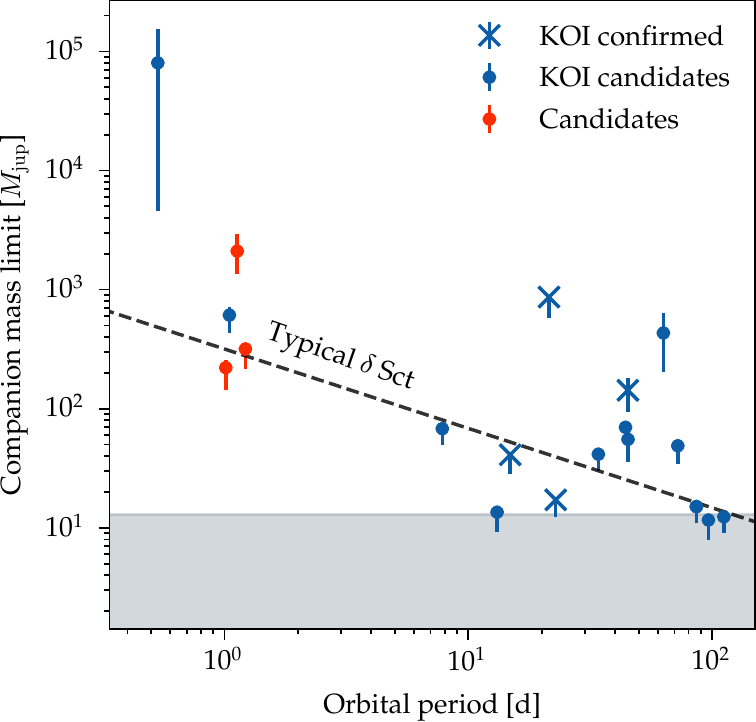}
    \caption{Results of pulsation timing on the stars in our sample. The dashed black line indicates the expected sensitivity for a typical \kepler \dsct \citep{Hey2020Forward}. The \rev{gray} shaded region denotes the area of brown dwarf masses.}
    \label{fig:ptiming}
\end{figure}

\input{tables/pm_results}

We fitted each light curve with the model in Eq.~\ref{eq:luminosity}, using a Hamiltonian Markov Chain Monte Carlo No-U-Turn (NUTS) sampler implemented in \textsc{PyMC3} \citep{Salvatier2016Probabilistic}. The prior on $a\sin{i}$ was chosen to be a bounded flat prior, ensuring that its value could not go negative. We fixed the eccentricity to be 0 for simplicity. The key aspect of this fit was that the orbital period was fixed to its value determined from transit analysis. The remaining priors are described in Table~\ref{tab:priors}.

We ran the model for 2000 tuning steps and 2000 draw steps over 2 chains simultaneously, resulting in a posterior size of 4000 points. We summarize the results of the pulsation timing model for each system in Table~\ref{tab:pm_results}, and plot the maximum possible companion masses in Fig~\ref{fig:ptiming}. We report the mass function ($f$), calculated from the value of a$\sin{i}$:

\begin{equation}
    f(m_1, m_2 \sin{i}) = \frac{4 \pi^2 c^3}{G P_{\rm orb}^2} \Big(a\sin{i} \Big)^3,
\end{equation}

\noindent where $c$ and $G$ are the speed of light and gravitational constant, respectively. The RV semi-amplitude ($K_1$) follows from the mass function,

\begin{equation}
    K_1 = (2 \pi G)^{1/3} \sqrt{1 - e^2} \Big[\frac{f(m_1, m_2 \sin{i})}{P_{\rm orb}} \Big] ^{1/3},
\end{equation}

\noindent which provides an estimate of the RV precision necessary to observe the planetary companion, if it exists. We note that the posterior distribution of nearly all of the stars for the a$\sin{i}$ and related mass quantities is highly asymmetric. This is a result of the nature of the sampling: although the distribution for most of the systems is centered around 0, the prior on $a\sin{i}$ ensures that the mass cannot go negative. Thus, the constraint on the mass is largely provided by how narrow the distribution is. We therefore report the 95th percentile of each posterior distributions, which is the value that encompass 2$\sigma$ of the posterior distribution.

For our sample, the pulsation timing model sets an upper limit on the companion mass limit that ranges from 33 to 90,000~M$_{\rm Jup}$. Most of these mass limits are too large to be useful because of the short orbital periods, which strongly limits the sensitivity of pulsation timing. However, most of the longer period KOI candidates are very well constrained. \rev{\citet{Hey2020Forward} found that a typical \dsct variable in the \kepler\ LC data had a pulsation timing precision of around 2~s, corresponding to the black dashed line in Fig.~\ref{fig:ptiming}. Stars that fall above this line are either of low SNR, or potentially have stellar mass companions that would cause a non-zero signal. Stars below this line are of higher SNR, and are thus provide better limits on the mass of their companion.} For the best case, KIC~6032370, the 95th percentile of the mass limit is less than 33~$M_{\rm Jup}$, implying a necessary RV precision of at least 1.0~km/s. Obtaining such precise RVs for a \dsct variable would be highly challenging, although significant work has gone into ameliorating the effects of rotation and pulsations on RV extraction in A/F stars \citep{Galland2005Extrasolara,Galland2006Extrasolar, Lagrange2009Extrasolarb,Desort2007Planets}.

A downside to this method is that it cannot decisively rule out candidates as false positives. This is because some companion mass limits lie well above the typical \dsct sensitivity line, even though they are confirmed planets. The reason for this is that the \dsct oscillations are either not coherent enough, or the SNR of the pulsations is too low \citep{Compton2016Binary}. Despite this, it is still useful in providing limits on possible future surveys.

\section{Discussion}

\begin{figure}
    \centering
    \includegraphics{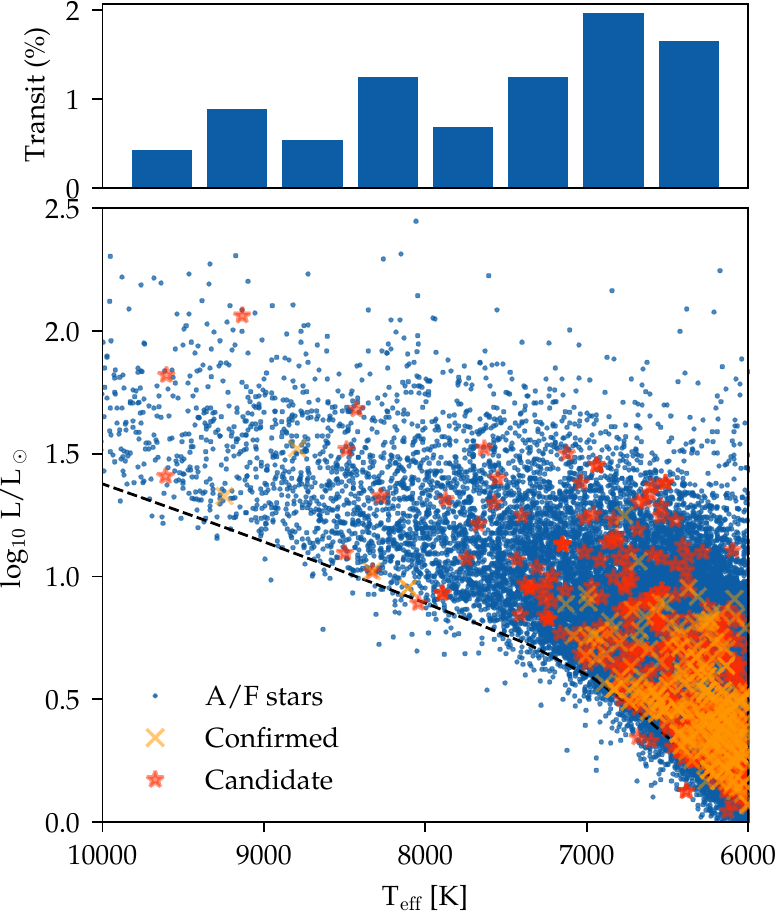}
    \caption{HR diagram of the A/F stars (blue) with the confirmed (orange crosses) and candidate (red stars) overlaid. The black line marks the zero age main sequence for Fe/H = 0.1 dex. The top panel shows the combined (confirmed \& candidate) occurrence rate as a function of temperature.}
    \label{fig:occurrence}
\end{figure}

Fig.~\ref{fig:occurrence} shows the occurrence rate of transiting planets among the hot A/F stars (6500~K to 10000~K) by using the catalog of A/F stars from \citet{Murphy2019Gaiaderived}, and planetary candidates from the KOI with temperatures and luminosities provided from \citet{Berger2020GaiaKepler}.

We calculated the occurrence rate as a function of temperature in 500~K bins by dividing the combined candidate and confirmed planetary systems by the number of A/F stars in each temperature bin. With such a small sample size of transiting planets above 7000~K, we cannot reliably provide any conclusions on the occurrence rate. Indeed, for stars between 7500 and 8500~K there are few confirmed planets, and only nine candidate planets between 8000 and 8500~K. While we cannot draw conclusions from such a small sample size, we suggest that the dearth of planetary candidates could be due to the fact that the \dsct occurrence rate peaks at around 8000~K \citep{Murphy2019Gaiaderived}. If so, there may still be a significant number of transits in the \kepler data hidden by the pulsations that our pipeline was not successful in identifying.


Note that we can only comment on the occurrence rate for transiting planets. The KOI list does not contain candidate planets which have been detected by other methods, and thus, they have been left out. The most obvious example of this is the single planet discovered by pulsation timing around a \dsct variable: KIC~7917485 \citep{Murphy2016Planet}.

If we assume that all of our candidates in Section~\ref{sec:results} are false positives, then we can estimate an upper limit for the occurrence rate ($\epsilon$) based on our transiting pipeline. The number of candidates $n$ from $N$ observations with a true occurrence rate $\epsilon$ is distributed according to the binomial distribution

\begin{equation}
    B(n|N, \epsilon) = \frac{N!}{n!(N-n)!}\epsilon^n(1-\epsilon)^{N-n},
\end{equation}

\noindent $N$=2354 is the number of \dsct variables in our sample, and $n$=0 is the number of transiting systems. By using Bayes' theorem, the posterior for $\epsilon$ is given by the beta distribution:

\begin{equation}
    \epsilon = B(n+1, N+1).
    \label{eqn:beta}
\end{equation}

\noindent Evaluating this distribution at the 95th percentile gives an upper limit on the occurrence rate of 0.13\% for our pipeline, which was tested to have a 100\% recovery rate on planets larger than 0.5~R$_{\rm Jup}$ with orbital periods of 10 days, falling to 50\% at 0.25~R$_{\rm Jup}$ (see Sec.~\ref{transit_search}).

A detailed estimate of the number of transiting planets around A/F stars expected in \textit{TESS} is beyond the scope of this paper. However, we can make a rough estimate based on the number of \dsct variables in \textit{Kepler} (2354) and the number of late A/F-type stars between 6,500 and 10,000~K (12,135). Dividing these numbers, we estimate the \dsct occurrence rate to be 19.3\%. \textit{TESS} has observed 45,232 unique stars in the same temperature range up to Sector 30. Thus, we expect \textit{TESS} to already have light curves for approximately 9000 \dsct stars. 

We used a Monte-Carlo approach to propagate uncertainties and estimate the expected number of planet-hosting \dsct stars by randomly drawing 50,000 samples from our posterior for $\epsilon$ (Eq.~\ref{eqn:beta}). We then evaluated distribution for $n$ for each sample from $B(n|N, \epsilon)$, and obtained an upper limit (95th percentile) of 12 planets. That is, given the values above and the efficiency of our pipeline, we predict that at the 95th percentile there are 12 detectable planets above 0.5~R$_{\rm Jup}$ in the current \textit{TESS} 2-minute cadence data. At the 50th percentile, there are 2 expected detectable planets.

\begin{figure}
    \centering
    \includegraphics{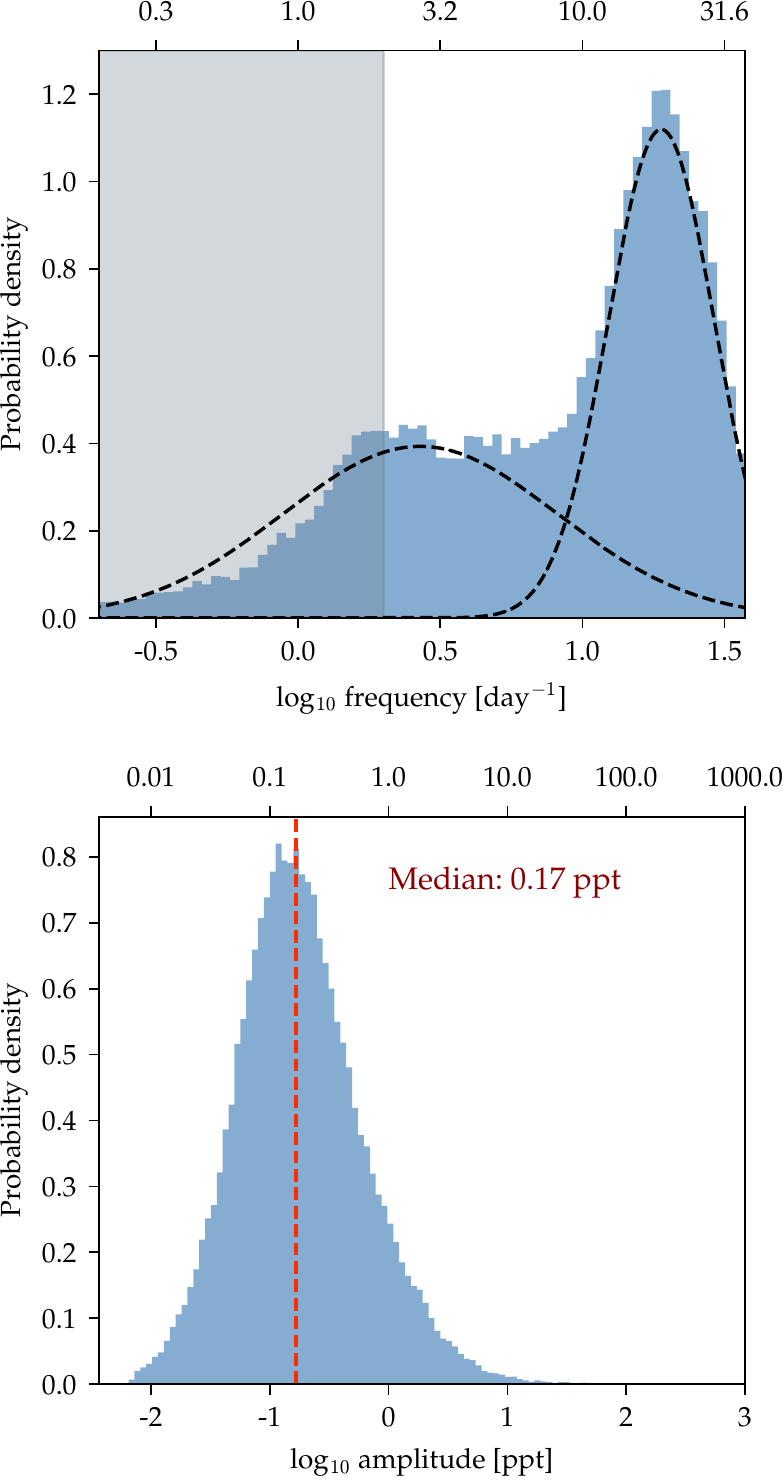}
    \caption{Frequency (top) and amplitude (bottom) distributions for all \kepler \dsct stars in our sample. The dashed black lines in the frequency distribution are the modeled components of the simple Gaussian Mixture Model, showing the two frequency regimes present in \dsct stars: high frequency p-modes and low-frequency g-modes. \rev{The gray region denotes the typical range of orbital frequencies. In the amplitude distribution, the median is marked in red. The top x-axis on both plots is the non-log version of the x-axis.}}
    \label{fig:dsct_dist}
\end{figure}

Another outcome of our cleaning routine is the frequency, amplitudes, and phases of almost every \dsct pulsation down to 4 SNR in the \kepler\ sample. This results in 84,352 values across all the stars in our sample, which represents the most complete set of pulsation data for the \dsct stars ever compiled. In the context of exoplanet analysis, this allows us to explore the parameter space for transits. We show in Fig.~\ref{fig:dsct_dist} the complete frequency and amplitude distribution of all stars in our sample. The frequency distribution is clearly split into two distinct distributions. The low-frequency region is dominated by g-modes, with some contamination from rotational variability. The high-frequency region is the \dsct p-modes.

We modeled both frequency distributions as a simple Gaussian Mixture Model (GMM) assuming two mixed distributions. Overlaid on Fig.~\ref{fig:dsct_dist} is the frequency range of planetary transits, which overlaps with the peak of g-mode pulsations. This is why it is useful to have a metric to distinguish low-frequency intrinsic signals from transit signals. In our case, the BIC worked remarkably well. 

The distribution of amplitudes peaks at a median value of 166~ppm. Although most of the high amplitudes in this distribution are due to p-mode pulsations outside the frequency range of transit signals, the transit will nevertheless be ``hidden'' beneath such a high noise floor. We advise that all such high-frequency signals be removed prior to searching for transits in the \dsct variables.

\begin{figure}
    \centering
    \includegraphics{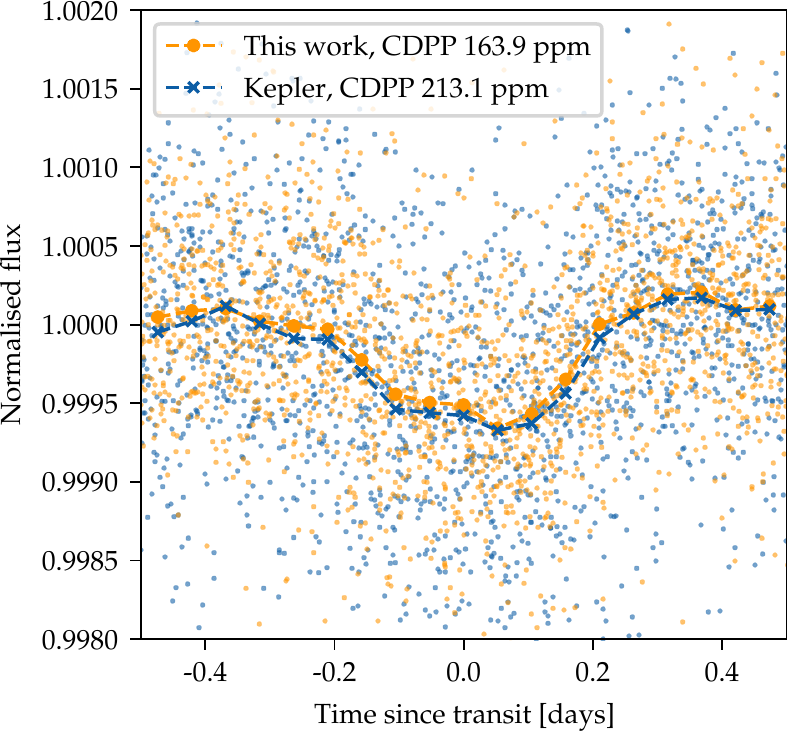}
    \caption{Folded light curves of KIC~9289704 for the cleaned light curves created in this work (orange), and those made by the \textit{Kepler} validation pipeline (blue). The dashed lines show the binned flux of the transit.}
    \label{fig:cdpp_compare}
\end{figure}

Finally, we performed a quick comparison of our cleaned light curves against the automated \textit{Kepler} planet search pipeline \citep{Jenkins2017Kepler, Twicken2018Kepler}. We selected one of our low-SNR candidates that is listed as a KOI and found by our pipeline, KIC~9289704, and calculated the combined differential photometric precision (CDPP) for our cleaned light curve, and the de-trended \textit{Kepler} pipeline light curve used in the planet validation reports. 
The CDPP is a metric that estimates the amount of remaining scatter after all long term trends have been removed. We calculated the CDPP using \textsc{Lightkurve}, at a fixed transit duration of 6.5 hours, and found that our pipeline performed somewhat better with a CDPP of 163.9~ppm, compared to the \textit{Kepler} CDPP of 213.1~ppm (Fig.~\ref{fig:cdpp_compare}). In the \textit{Kepler} pipeline, the light curves are corrected by process termed `whitening', whereby a wavelet-based matched filter is used to correct over a range of band passes. This has the effect of removing pulsations but does so in a less targeted way than our approach, hence the difference in performance.

\section{Conclusion}
In this paper, we searched for transit events among the \dsct variables by subtracting their pulsations through an automated routine. We identified 3 possible new candidates, but warn that they are likely false positives, with shallow transits that will be difficult to confirm photometrically without \textit{Kepler}. We also identified 13 additional Kepler Objects of Interest which show \dsct pulsations, and used these pulsations to constrain the mass of the planetary companion.

Despite the dearth of planetary candidates found around the A/F stars, we note that the pipeline we presented is applicable to all space-based photometric missions. Both \textit{TESS} and \textit{PLATO} are expected to yield a massive amount of \dsct variables across the sky. Because of this, techniques for dealing with coherent pulsations in light curves will be fundamental for searching for planets around the hot variable stars.

\acknowledgments

\rev{We thank the anonymous referee for their careful review, which greatly improved the quality of this manuscript.} DRH gratefully acknowledges the support of the AGRTP scholarship.  We also acknowledge support from the Australian Research Council (DP210103119) and from the Danish National Research Foundation (Grant DNRF106) through its funding for the Stellar Astrophysics Center (SAC). 
We acknowledge the traditional owners of the lands on which the University of Sydney is located, the Gadigal people of the Eora Nation, and we pay our respect to the knowledge embedded forever within the Aboriginal Custodianship of Country. 
BJSP would like to acknowledge the traditional owners of the land on which the University of Queensland is situated, the Turrbal and Jagera people. We pay respects to their Ancestors and descendants, who continue cultural and spiritual connections to Country.
This paper includes data collected by the \kepler mission and obtained from the MAST data archive at the Space Telescope Science Institute (STScI). Funding for the \kepler mission is provided by the NASA Science Mission Directorate. STScI is operated by the Association of Universities for Research in Astronomy, Inc., under NASA contract NAS 5–26555.
This research has made use of the NASA Exoplanet Archive, which is operated by the California Institute of Technology, under contract with the National Aeronautics and Space Administration under the Exoplanet Exploration Program.

\vspace{5mm}
\software{astropy \citep{astropy:2013, astropy:2018},  
 lightkurve \citep{lightkurve, GeertBarentsen2019KeplerGO},
 PyMC3 \citep{Salvatier2016Probabilistic},
 TLS \citep{Hippke2019Optimized},
matplotlib \citep{Hunter2007Matplotlib},
numpy \citep{Oliphant2015Guide}.
          }


\bibliography{library}{}
\bibliographystyle{aasjournal}

\end{document}

%% file: commands.tex
\newcommand\kepler{\emph{Kepler}\,}

\newcommand{\dsct}{\mbox{$\delta$~Sct} }

%% file: authors.tex
\author[0000-0003-3244-5357]{Daniel R. Hey}
\affil{School of Physics, Sydney Institute for Astronomy (SIfA) \\
The University of Sydney, NSW 2006, Australia}
\affil{Stellar Astrophysics Centre, Department of Physics and Astronomy \\
Aarhus University, DK-8000 Aarhus C, Denmark}

\newcommand{\unsw}{School of Physics, University of New South Wales, Sydney, NSW 2052, Australia}
\newcommand{\udash}{UNSW Data Science Hub, University of New South Wales, Sydney, NSW 2052, Australia}

\author[0000-0001-7516-8308]{Benjamin~T.~Montet}
\affil{\unsw}
\affil{\udash}

\author[0000-0003-2595-9114]{Benjamin~J.S.~Pope}
\affil{School of Mathematics and Physics, The University of Queensland, St Lucia, QLD 4072, Australia}
\affil{Centre for Astrophysics, University of Southern Queensland, West Street, Toowoomba, QLD 4350, Australia}

\author[0000-0002-5648-3107]{Simon J. Murphy}
\affil{School of Physics, Sydney Institute for Astronomy (SIfA) \\
The University of Sydney, NSW 2006, Australia}
\affil{Stellar Astrophysics Centre, Department of Physics and Astronomy \\
Aarhus University, DK-8000 Aarhus C, Denmark}

\author[0000-0001-5222-4661]{Timothy R. Bedding}
\affil{School of Physics, Sydney Institute for Astronomy (SIfA) \\
The University of Sydney, NSW 2006, Australia}
\affil{Stellar Astrophysics Centre, Department of Physics and Astronomy \\
Aarhus University, DK-8000 Aarhus C, Denmark}


%% file: tables/results.tex
\begin{table*}
    \centering
    \begin{tabular}{lccccccc}
    \hline
        KIC  &  Cadence & Transit period  &   Transit $T_0$ & Transit duration & Radius & a/R$_*$ & $\frac{\rho_{\rm implied}}{\rho_{\rm calculated}}$ \\
            &  &  (d) &   (d) &   (d)  & (R$_\oplus$)  \\
    \hline
    New candidates \\
    \hline
    7767699 & LC & $1.12828^{+0.00003}_{-0.00003}$ & $121.66^{+0.02}_{-0.02}$ & $0.16^{+0.02}_{-0.02}$ & $0.56^{+0.09}_{-0.07}$ & $0.98^{+0.2}_{-0.2}$ & $0.53$\\
    8249829 & LC & $1.0127^{+0.0001}_{-0.0001}$ & $1472.679^{+0.006}_{-0.009}$ & $0.090^{+0.009}_{-0.008}$ & $2.8^{+0.4}_{-0.3}$ & $1.6^{+0.3}_{-0.5}$ & $2.6$\\
    9895543 & LC & $1.21889^{+0.00002}_{-0.00002}$ & $121.42^{+0.02}_{-0.01}$ & $0.15^{+0.02}_{-0.02}$ & $1.0^{+0.1}_{-0.1}$ & $1.1^{+0.2}_{-0.3}$ & $0.81$\\
    \hline
    KOI planet candidates \\ with \dsct pulsations \\
    \hline
3964109 & SC & $21.41624^{+0.00007}_{-0.00007}$      & $176.250^{+0.003}_{-0.003}$ & $0.295^{+0.004}_{-0.004}$ & $2.1^{+0.4}_{-0.2}$ & $23.4^{+0.6}_{-0.6}$ & $0.86$ \\ 
 9111849 & LC & $63.0732^{+0.0003}_{-0.0003}$        & $191.253^{+0.003}_{-0.004}$ & $0.225^{+0.005}_{-0.005}$ & $17.8^{+4.5}_{-8.9}$ & $62.8^{+25.2}_{-25.2}$ & $3.77$ \\ 
 \textbf{\textit{9845898}} & LC & $45.1262^{+0.0005}_{-0.0005}$ & $165.421^{+0.009}_{-0.01}$ & $0.28^{+0.01}_{-0.01}$ & $2.3^{+0.8}_{-0.3}$ & $46.2^{+17.0}_{-17.0}$ & $0.96$ \\ 
 \textbf{\textit{5202905b}} & LC & $22.8245^{+0.0002}_{-0.0002}$ & $144.468^{+0.008}_{-0.009}$ & $0.28^{+0.01}_{-0.01}$ & $3.4^{+0.6}_{-0.7}$ & $25.8^{+4.9}_{-4.9}$ & $0.94$ \\ 
 \textbf{\textit{5202905c}} & LC & $14.8445^{+0.0002}_{-0.0002}$ & $143.57^{+0.01}_{-0.01}$ & $0.25^{+0.02}_{-0.02}$ & $2.5^{+0.4}_{-0.5}$ & $18.7^{+7.2}_{-7.2}$ & $0.84$ \\ 
 {6116172b} &LC &  $69.64679$              & $138.28306$      & $0.04^{+0.02}_{-0.02}$ & $2121^{+1290}_{-467}$ & $799^{+232}_{-232}$ & $4968$ \\ 
 {6116172c} & LC & $111.6555^{+0.0007}_{-0.0007}$      & $148.894^{+0.005}_{-0.005}$ & $0.04^{+0.02}_{-0.02}$ & $2121^{+1290}_{-467}$ & $799^{+232}_{-232}$ & $4968$ \\ 
 {6116172d} &LC &  $86.13202$              & $164.22370 $          & $0.04^{+0.02}_{-0.02}$ & $2121^{+1290}_{-467}$ & $799^{+232}_{-232}$ & $4968$ \\ 
 {11013201b} &SC &  $13.11896^{+0.00001}_{-0.00001}$ & $261.5394^{+0.0003}_{-0.000}$ & $0.1794^{+0.0005}_{-0.0005}$ & $8.3^{+0.5}_{-2.5}$ & $23.69^{+0.04}_{-0.04}$ & $2.51$ \\ 
 {11013201c} &SC &  $7.82187^{+0.00002}_{-0.00002}$ & $123.499^{+0.003}_{-0.003}$ & $0.160^{+0.003}_{-0.003}$ & $2.8^{+0.2}_{-0.8}$ & $15.7^{+1}_{-1}$ & $2.04$ \\ 
 {6032730} & SC & $96.509^{+0.001}_{-0.001}$       & $223.808^{+0.008}_{-0.008}$ & $0.02^{+0.01}_{-0.01}$ & $860^{+222}_{-274}$ & $639^{+283}_{-283}$ & $1514$ \\ 
 {5617259} & LC & $1.04813$       & $132.35280 $           & $0.23628 $ & $4.5^{+0.4}_{-2.4}$ & $1.53664 $ & $0.42$ \\ 
 {9775385} &SC &  $45.163^{+0.005}_{-0.005}$ & $159.9^{+0.1}_{-0.1}$       & $0.30^{+0.09}_{-0.09}$ & $2.4^{+1.2}_{-0.4}$ & $48.6^{+18.6}_{-18.6}$ & $1.97$ \\ 
 {6670742} &SC &  $72.282^{+0.001}_{-0.001}$ & $156.75^{+0.02}_{-0.02}$      & $0.10^{+0.03}_{-0.03}$ & $13.1^{+2.0}_{-4.6}$ & $224.0^{+111.0}_{-111.0}$ & $211$ \\ 
 \textit{9289704} & LC & $34.136^{+0.002}_{-0.002}$ & $157.80^{+0.05}_{-0.05}$      & $0.38^{+0.08}_{-0.08}$ & $14.0^{+4.7}_{-8.8}$ & $29.1^{+10.0}_{-10.0}$ & $0.68$ \\ 
 \textit{3965201} & LC & $44.114^{+0.002}_{-0.002}$ & $164.83^{+0.03}_{-0.02}$      & $0.16^{+0.04}_{-0.04}$ & $4.9^{+0.9}_{-1.6}$ & $86.3^{+3.3}_{-3.3}$ & $56$ \\ 
 11180361 & SC &  $0.53306$  & $171.22744 $              & $0.07759 $ & $133.6^{+26.0}_{-78.0}$ & $1.45285 $ & $2.13$ \\ 
      \hline
    \end{tabular}
    \caption{Properties of the systems investigated in this paper. IDs in italics mean that the system was also found in our prewhitening transit search, and IDs in bold mean the system is confirmed. \rev{For the new candidates, uncertainties were obtained from the MCMC analysis performed in Section~\ref{sec:results}. The uncertainties for the KOI planet candidates were taken from the KOI catalog. Some systems in the KOI did not have uncertainties and thus they are left blank.} $b$, $c$, and $d$ designations in the ID denote identification for multiple systems.}
    \label{tab:planet}
\end{table*}


%% file: tables/fp.tex
\begin{table}
    \centering
    \begin{tabular}{ll}
    \hline
        KIC & Reason  \\
    \hline
    4380834 & a/R$_*$ << 1 \\
    5724523 & a/R$_*$ << 1 \\
    9471419 & a/R$_*$ << 1 \\
    9655419 & Did not converge \\
    8912730 & Did not converge \\
    7779942 & Did not converge \\
    7618364 & Did not converge \\
    5702637 & Did not converge \\
    5565497 & Did not converge \\
    5302643 & Did not converge \\
    4952341 & Did not converge \\
    4035667 & Did not converge \\
    3444098 & Did not converge \\
    2167444 & Did not converge \\
    11607193 &Did not converge \\
\hline
    \end{tabular}
    \caption{Table of false positives identified in the search and their reason for exclusion.}
    \label{tab:fp}
\end{table}

%% file: tables/pm_dists.tex
\begin{table}
    \centering
    \begin{tabular}{ll}
    \hline
    Prior   &   Description \\
    \hline  
    \textit{Pulsations} \\
    $\omega_j$:  $\mathcal{N} \sim (P, 5)$  & Angular frequency of oscillations\\
    $A_j$: $\mathcal{N} \sim (P, 5)$                                 & Amplitude of oscillation modes\\
    $\phi_j$: $\mathcal{N} \sim (P, 5)$            & Phase of oscillation modes\\
    \\
    \textit{Orbit model} \\
    $P_{\rm orb}$: $\mathcal{N} \sim (P^*, P_e^*)$ & Orbital period (days) \\
    $\rm{a_j}\sin{i}/c$: $\mathcal{F} $   &   Projected semi-major axis (s) \\
    $\varpi$: $\mathcal{U} \sim (0, 2\pi)$ & Ascending node and periapsis angle \\
    $\phi_p$: $\mathcal{U} \sim (0, 2\pi)$ &  Phase of periapsis passage\\
    \hline
    \end{tabular}
    \caption{Priors used in the model. Note that $\omega_j$, $A_j$ and $\phi_j$ are defined as a vector of length equal to the number of oscillation modes used in the model. A star on the value indicates that it has been obtained from transit information.}
    \label{tab:priors}
\end{table}

%% file: tables/pm_results.tex
\begin{table}
    \centering
    \begin{tabular}{lcccc}
    \hline
        KIC ID  & KOI & a$_1\sin{i}/c$  &  Mass limit & $K_{1, \rm{min}}$  \\
            &  & (s) &   ($M_{\rm Jup}$) & (km/s) \\
    \hline
    Candidates \\
    \hline
    7767699  & & $30$ &  $4300$ &  $580$ \\ 
    8249829  & & $4.7$ &  $670$ &  $100$ \\ 
    9895543  & & $6.9$ &  $920$ &  $120$ \\ 
    \hline
    Known \\
    \hline
    3964109  & K00393.01 & $150$ &  $2200$ &  $150$ \\ 
    9111849  & K02042.01 & $65$ &  $790$ &  $23$ \\ 
    \textbf{9845898}  & K02423.01 & $33$ &  $310$ &  $16$ \\ 
    \textbf{5202905b}  & K01932.01 &$2.8$ &  $48$ &  $2.7$ \\ 
    \textbf{5202905c}  &K01932.02& $4.7$ &  $110$ &  $6.9$ \\ 
    6116172b  & K06142.01 &$6.0$ &  $44$ &  $1.5$ \\ 
    6116172c  & K06142.02 &$5.7$ &  $35$ &  $1.1$ \\ 
    6116172d  & K06142.03 &$6.0$ &  $44$ &  $1.5$ \\ 
    11013201b  & K00972.01 &$1.2$ &  $37$ &  $2.0$ \\ 
    11013201c  & K00972.02 &$4.4$ &  $190$ &  $12$ \\
    6032730  & K06141.01 &$4.4$ &  $33$ &  $1.0$ \\ 
    5617259  & K07733.01 &$12$ &  $1900$ &  $250$ \\ 
    9775385  & K04462.01 & $10$ &  $120$ &  $5.0$ \\ 
    6670742  & K06753.01& $14$ &  $140$ &  $4.3$ \\ 
    9289704  & K02222.01 &$6.0$ &  $120$ &  $3.9$ \\ 
    3965201  & K07546.01 &$14$ &  $190$ &  $6.9$ \\ 
    11180361  & K00971.01 &$320$ &  $88000$ &  $13000$ \\ 
      \hline
    \end{tabular}
    \caption{Results of the pulsation timing model. Note that the values are reported as the 95th percentile of the posterior distribution. A bold ID indicates the planet is confirmed. For reference, 1~au/c$\sim$500~s.}
    \label{tab:pm_results}
\end{table}